\documentstyle[aps,pre,epsf]{revtex}

\begin{document}
\draft  
\title{Three-dimensional Ising model in the fixed-magnetization ensemble: \\
       a Monte Carlo study}
\author{H.~W.~J. Bl\"ote and J.~R.~Heringa}
\address{Faculty of Applied Physics, Delft University of Technology,
         P.O. Box 5046, 2600 GA Delft, The Netherlands}
\author{M.~M.~Tsypin}
\address{Department of Theoretical Physics, Lebedev Physical Institute,
         117924 Moscow, Russia}

\maketitle
\begin{abstract}
We study the three-dimensional Ising model at the critical point
in the fixed-magnetization ensemble, by means of the recently
developed geometric cluster Monte Carlo algorithm. We define a
magnetic-field-like quantity in terms of microscopic spin-up and
spin-down probabilities in a given configuration of neighbors.
In the thermodynamic limit, the relation between this field and
the magnetization reduces to the canonical relation $M(h)$. However,
for finite systems, the relation is different. We establish a
close connection between this relation and the probability
distribution of the magnetization of a finite-size system in the
canonical ensemble.
\end{abstract}
\pacs{05.50.+q, 64.60.Cn, 05.10.Ln, 75.40.Mg}

\section{Introduction}
\label{sec:intr}
Second order phase transitions display many interesting and subtle
properties associated with scale invariance and universality at
critical points. Some of these, such as power-law singularities of
the free energy and other quantities at the critical point, and their
critical exponents and amplitudes, have been studied rather thoroughly
(see, e.~g., \cite{ZJbook}).
Among less investigated items are the universal characteristics
of finite-size effects. These are important for analysis 
of experiments with finite samples, as well as for computer simulations,
which necessarily have to deal with finite-size systems.

In clear distinction from, for example, critical indices,
the finite size effects depend crucially on the nature of the 
statistical ensemble under consideration. To be concrete, 
let us consider one of the standard model systems of the 
phase transition theory --- the three-dimensional (3D) Ising
model on the simple cubic lattice, with nearest-neighbor
interactions.

According to universality, this model describes the critical properties
of a wide range of systems of a different physical nature,
including second order phase transitions in uniaxial
magnetic systems and the liquid-vapor critical point.
The statistical ensemble most commonly used (usually 
denoted as canonical ensemble) is defined by the partition
function
\begin{equation} \label{Zcanon}
  Z_c(h) = \sum_{ \{s_i \} } \exp \Bigl\{
  \beta \sum_{\langle ij \rangle} s_i s_j + h \sum_i s_i \Bigr\},
  \qquad s_i = \pm 1,
\end{equation}
where the sum includes all the $2^N$ possible configurations of a
total number of $N$ spins. This ensemble is perfectly
natural for applications to magnetic phase transitions, where
$s_i = \pm 1$ corresponds to physical spin at the lattice site $i$ 
pointing up or down, respectively. However, in the language of the
lattice gas, $s_i = \pm 1$ corresponds to an occupied or unoccupied
site, respectively: the total number of particles fluctuates. The
canonical partition sum Eq.~(\ref{Zcanon}) in the spin language thus
corresponds with the grand partition sum in the lattice gas language.
To avoid confusion, we emphasize that in the rest of this paper we
are using the word `canonical' in the language of the Ising model:
it means that the magnetization is allowed to fluctuate freely.

In contrast, many real experiments and simulations of liquid-vapor
systems are performed with a fixed number of particles. Fixing the
number of lattice-gas particles is equivalent, in the language of
magnetic systems, to fixing the total magnetization
$M_{\rm total} \equiv \sum_{i=1}^N s_i$ of the system, or,
in other words, fixing the average magnetization per spin
$M \equiv {1\over N}M_{\rm total} = {1\over N} \sum_{i=1}^N s_i$.

Thus we will be interested in the properties of the 3D Ising model 
in the fixed-magnetization ensemble,
\begin{equation} \label{ZfixedM}
  Z_f(M) = \sum_{ \{s_i \}: \sum_i s_i = NM } \exp \Bigl\{
  \beta \sum_{\langle ij \rangle} s_i s_j \Bigr\} = 
  \sum_{ \{s_i \} } \delta_{NM,\sum_i s_i} 
  \exp \Bigl\{ 
  \beta \sum_{\langle ij \rangle} s_i s_j \Bigr\}.
\end{equation}
Note that the magnetic field $h$ is absent; it can only contribute
a constant factor.

One of the main difficulties encountered in computer simulation studies 
of critical phenomena is the critical-slowing-down phenomenon. For a
number of spin models
in the canonical ensemble this problem has been largely overcome
with the invention of cluster algorithms \cite{SwWa,Wolff}.
Until recently, no similarly useful algorithm has become available
for the fixed-magnetization ensemble. This situation has now changed 
by the development of a geometric cluster algorithm \cite{HB98a,HB98b}.
We have used this algorithm extensively in this work to efficiently
simulate systems of fixed magnetization at the critical point.

\begin{figure}

\centerline{
\epsfxsize=16.5cm
\epsfbox{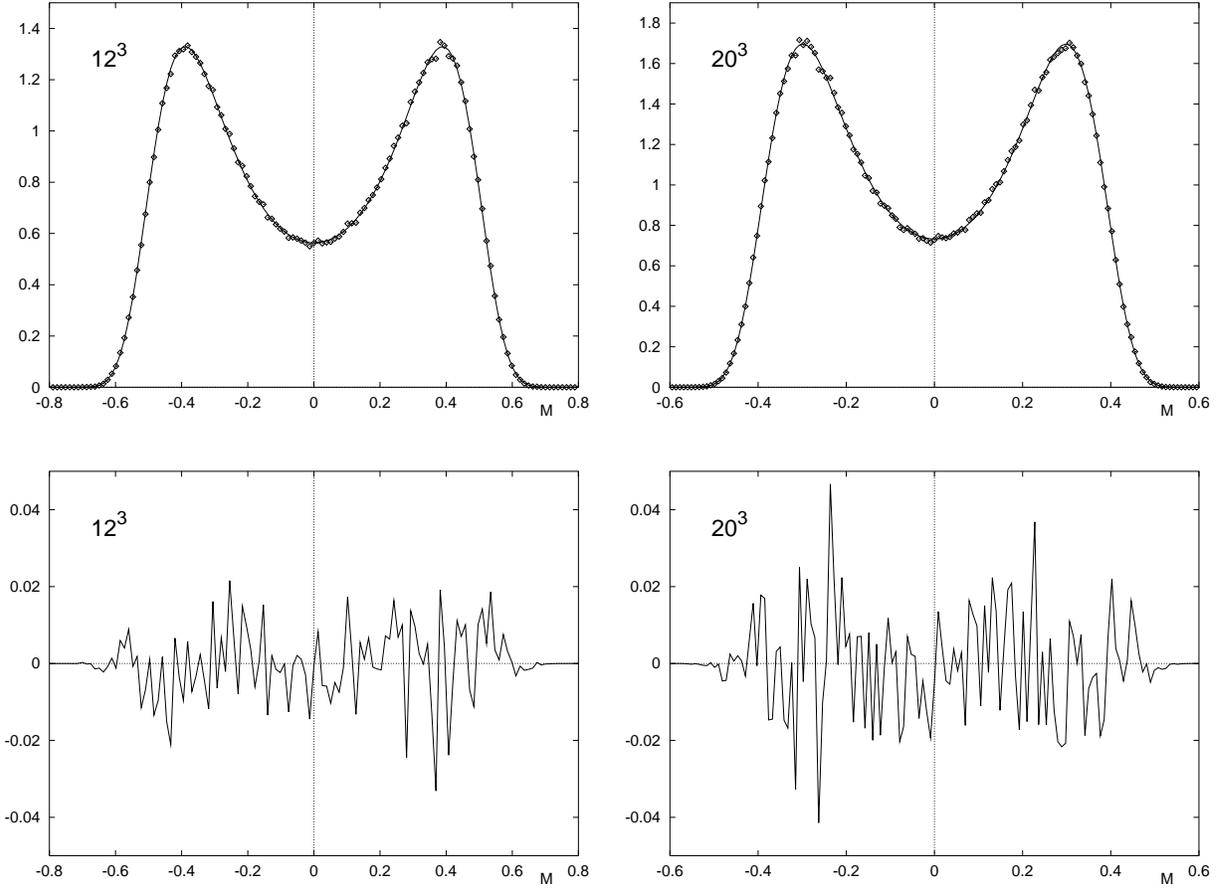}
\bigskip
}

\caption{
Top: Probability distributions $P(M)$ of the magnetization per spin
$M = {1\over N} \sum_{i=1}^N s_i$, at the critical point
($\beta=0.221654$, $h=0$), for the 3D Ising model,
Eq.~(\protect\ref{Zcanon}), in a cubic box with periodic
boundary conditions, for two lattice sizes: $12^3$ (left)
and $20^3$ (right). The Monte Carlo data were obtained
using the Swendsen-Wang cluster algorithm (720000 configurations
for each lattice size). The solid line is a fit according to
Eq.~(\protect\ref{PMansatz}).
For the $12^3$ lattice, $a = 0.268(13)$, $c = 0.859(8)$, $M_0=0.3892(11)$.
For the $20^3$ lattice, $a = 0.209(9)$, $c = 0.839(11)$, $M_0=0.2984(9)$.
Bottom: the difference between the data and the fit.
}
\label{fig1}
\end{figure}

In the canonical Ising model described by Eq.~(\ref{Zcanon}), the
magnetic field $h$ is an adjustable external parameter. In contrast,
the magnetization is a fluctuating observable. For each configuration
taken from the ensemble one can sample its magnetization
per spin $M = {1\over N} \sum_{i=1}^N s_i$. Having accumulated $M$ over
a sufficiently large set of configurations, one can construct the
probability distribution $P(M)$ 
\cite{Binder81,HilferW,Stauffer98,Kaneda99,TsyBlo}, 
and determine various expectation values such as $\langle M \rangle$,
$\langle M^2 \rangle$. Examples of such probability
distributions at the critical point are shown in Fig. \ref{fig1}.

On the other hand, for systems in the fixed-magnetization ensemble
described by Eq.~(\ref{ZfixedM}), the roles of $h$ and $M$ 
are interchanged: now $M$ is the adjustable parameter, and it is
intuitively clear that there should be some way to define an
observable, which we denote by $\tilde{h}$ (to avoid confusion
with $h$ in Eq.~(\ref{Zcanon})), that will correspond to the
magnetic field. Thus $\tilde{h}$ will be a fluctuating quantity,
that can be sampled on a microscopic level from configurations taken
from the fixed-magnetization ensemble. In the limit $N \to \infty$
in both ensembles (such that the correlation length vanishes in
comparison with the system size),
the fluctuations in $M$ and $\tilde{h}$ become negligible, and the
difference between $h$ and $\tilde{h}$ vanishes.

In Section \ref{sec:defh} we discuss the definition of $\tilde{h}$
and its properties. In Section \ref{sec:h-and-P} we establish the 
relation between the function $\tilde{h}(M)$ in the fixed-$M$
ensemble and the probability distribution $P(M)$ in the canonical
ensemble. We conclude with a discussion of the relation between
the function $\tilde{h}(M)$ in the fixed-$M$ ensemble and the
function ${h}(M)$ in the canonical ensemble, and with a summary
of our main results.

\section{The magnetic observable $\tilde{h}(M)$ for the fixed-$M$
ensemble}
\label{sec:defh}

We will now describe a definition of $\tilde{h}(M)$ that is based on
statistical analysis of the local environment of a given spin. 
By local environment we mean the set of neighbors with which this 
spin interacts. In our particular case of the Ising model with 
nearest-neighbor interactions on the simple cubic lattice,
the local environment consists of 6 spins on the neighboring sites.
The local environment has $2^6$ possible configurations which divide
in 7 types: 0+6$-$ (zero spins up, 6 down), 1+5$-$, 2+4$-$, 3+3$-$,
4+2$-$, 5+1$-$, 6+0$-$ (six spins up, zero down). The simplest way to 
Monte Carlo sample $\tilde{h}(M)$ is on the basis of the symmetric
case: 3+3$-$ \cite{HB98b}. For every Monte Carlo configuration, 
go through all sites, and select all spins with the required 3+3$-$
local environment. Then compute the average $\langle s_0 \rangle$ of
the selected spins, and define $\tilde{h}$ by
\begin{equation}
  \langle s_0 \rangle = \tanh \tilde{h} .
\label{defH33}
\end{equation}
It is also possible to employ, instead of 3+3$-$, any other of the 7
types of local environment. In these non-symmetric cases the
definition reads
\begin{equation}
  \langle s_0 \rangle = 
  \tanh \Bigl( \tilde{h} + \beta \sum_{i=1}^6 s_{0i} \Bigr) ,
\label{defH}
\end{equation}
where the $s_{0i}$ are the nearest neighbors of $s_0$.

One easily notices that the definition is constructed in such a way
that $\tilde{h}$ corresponds, on the mean field level, to the
external field $h$ in Eq.~(\ref{Zcanon}).

Now it is interesting to see what are the results of Monte Carlo
simulations for $\tilde{h}(M)$. As has already been demonstrated
in \cite{HB98a,HB98b}, at the critical temperature $\tilde{h}(M)$
practically coincides with the relation $h(M)$ in the canonical
ensemble as obtained by Monte Carlo simulations, provided $M$ is
sufficiently large, so that the correlation length is sufficiently
small in comparison with the system size and the finite-size effects
are suppressed. At the same time, the striking feature of $\tilde{h}(M)$
for not-so-large $M$ is its nonmonotonic behavior. First $\tilde{h}(M)$
goes negative at small $M$, then it begins to grow, and finally assumes
the usual behavior at larger $M$ \cite{HB98b}. This is clearly seen in
Fig.~\ref{fig2} (diamonds), which shows Monte Carlo results obtained
by means of the the geometric cluster algorithm \cite{HB98a,HB98b}.

In the remaining part of the paper, we will give the explanation of this
behavior (which turns out to be a peculiar kind of finite size effect),
by establishing a close relation between $\tilde{h}(M)$ in the fixed-$M$
ensemble, and the probability distribution $P(M)$ in the canonical
ensemble.

\section{
Connection between $\tilde{h}(M)$ in the fixed-$M$ ensemble
and the probability distribution of $M$ in the canonical ensemble
}
\label{sec:h-and-P}

Considering the fixed-$M$ ensemble, Eq.~(\ref{ZfixedM}), one notices 
that it can be obtained by taking the canonical ensemble (\ref{Zcanon}),
and cutting from it the subset satisfying the constraint, $\sum_i s_i = NM$.
Within this subset we still have the usual Boltzmann probabilities
$\exp\{ \beta \sum_{\langle ij \rangle} s_i s_j \}$ for individual 
configurations.

This makes it possible to establish a relation between $\tilde{h}(M)$ 
in the fixed-$M$ ensemble, and the properties of the system in the 
canonical ensemble. The definition of $\tilde{h}(M)$ described in 
Sect.~\ref{sec:defh} is equivalent to the following. Let us take
the fixed-$M$ ensemble and concentrate our attention on one particular 
lattice site, and on the spin located there. Let us perform the following
measurement. For every configuration consider the local environment
of our selected site. If it is not 3+3$-$, do not measure anything
for this configuration. If it is 3+3$-$, measure the selected spin 
$s_0$ and store it.
Finally, find $\langle s_0 \rangle$, and use Eq.~(\ref{defH33})
to determine $\tilde{h}$.

One notices that, as long as we are performing a thought experiment,
we need not care about the Monte Carlo statistics. We can just
stick to one site and get the same $\langle s_0 \rangle$ without
averaging over all sites, because they are equivalent.

Up to now we have distinguished between 7 types of local
environments, such as 3+3$-$. Let us go a bit further and 
treat separately all $2^6 $ possible local environments.
In other words, the measurement of $\langle s_0 \rangle$ is now
performed in an even smaller subset of the fixed-$M$ ensemble: also
the six spins forming the local environment of $s_0$ are fixed.
In the case that the predetermined local environment is of the
3+3$-$ type, it may seem that there is no interaction between $s_0$
and the remaining system of $N-7$ spins. Nevertheless, the fixed-$M$
ensemble probabilities that $s_0$ is $+1$ or $-1$, which we denote,
respectively, by $P_+$ and $P_-$, are not equal in general. These
probabilities may still depend
on the magnetization of the remaining system, which is coupled to
$s_0$ by the overall magnetization constraint:
\begin{equation}
 s_0 + \sum_{i=1}^6 s_{0i} + \sum_{i \in RS} s_i = \sum_{i=1}^N s_i 
 \equiv NM .
\end{equation}
The total magnetization of the system is thus expressed as the sum
of three terms: the local spin $s_0$, the sum of its six neighbors
and the magnetization of the remaining $N-7$ spins denoted as
$\sum_{i \in RS} s_i$, where $RS$ stand for ``remaining system''.

The conditional probabilities $P_\pm$ can be more explicitly written as
\begin{equation}
P_\pm = P(s_0=\pm 1 |NM,s_{01}\cdots s_{06})
\end{equation}
The two conditional arguments specify the total magnetization $NM$ and 
the states of the 6 neighbor spins. We now make the connection with
the canonical probabilities $P_c$ which include the magnetization as an
unconditional argument. We use the zero-field canonical probabilities
i.e., $h=0$ in Eq.~(\ref{Zcanon}).
\begin{equation}
P_\pm = P_c^{-1}(NM | s_{01}\cdots s_{06})
P_c(s_0=\pm 1,NM | s_{01}\cdots s_{06})
\end{equation}
We may slightly rewrite this by substitution of the probability $P_c$
by $\hat{P}_c$ which is equal but uses the magnetization of the
$N-7$ remaining spins as its second argument:
\begin{equation}
P_\pm = P_c^{-1}(NM | s_{01}\cdots s_{06}) \hat{P}_c(s_0=\pm 1,
\sum_{i \in RS} s_i=NM-\sum_{i=1}^6 s_{0i}-s_0 | s_{01}\cdots s_{06})
\end{equation}
Let us first consider the simplest case $\sum_{i=1}^6 s_{0i} = 0$. Thus
the canonical probability $\hat{P}_c$ does not depend on its first
argument, which can thus be skipped:
\begin{equation}
P_\pm = \frac{1}{2} P_c^{-1}(NM | s_{01}\cdots s_{06}) 
\hat{P}_c(\sum_{i \in RS} s_i = NM \mp 1 | s_{01}\cdots s_{06})
\end{equation}
Therefore,
\begin{equation}
 {P_+ \over P_-} =
 { \hat{P}_c ( \sum_{i \in RS} s_i = NM -1 | s_{01}\cdots s_{06}) \over  
   \hat{P}_c ( \sum_{i \in RS} s_i = NM +1 | s_{01}\cdots s_{06}) } .
\label{p+over_p-}
\end{equation}
The condition $s_{01}\cdots s_{06}$ in effect introduces a defect in
the remaining system: an octahedron-shaped bubble with six spins at
its vertices fixed, while the spin $s_0$ in the middle is decoupled and
plays no role any more.

Obviously, the ratio (\ref{p+over_p-}) could be obtained by performing 
a usual canonical ensemble simulation of such a system with a defect,
and measuring the probability distribution for its overall magnetization
$\sum_{i \in RS} s_i$. The value of the ratio (\ref{p+over_p-}) would 
then be given by the ratio of the heights of the two neighboring bins
in the corresponding histogram.

In all cases of practical interest for the study of the scaling limit 
(sufficiently large systems, sufficiently small magnetization)
the ratio (\ref{p+over_p-}) is close to 1. Otherwise a difference 
of one unit in the total magnetization would lead to a large change
of probability: this would obviously be far from the scaling limit.
Thus we always work with $\tilde{h} \ll 1$.

It is convenient to introduce a shorter notation
$P_{RS}(x) \equiv \hat{P}_c(\sum_{i \in RS}s_i=Nx|s_{01}\cdots s_{06})$
where $x$ may be read as the magnetization of the remaining system if the
factor $N$ instead of $N-7$ makes a negligible difference, i.e. for large
systems. We have to keep in mind that the notation $P_{RS}$ refers to a
system with a defect whose type is not explicitly shown.
Again restricting ourselves to local environments of the type
3+3$-$, we obtain
\begin{equation}
 \langle s_0 \rangle = 
 { P_+ - P_- \over P_+ + P_-} =
 {
 P_{RS}(M-{1\over N}) - P_{RS}(M+{1\over N})
 \over
 P_{RS}(M-{1\over N}) + P_{RS}(M+{1\over N})
 } .
\label{RS1}
\end{equation}
Thus we arrive at
\begin{equation}
 \langle s_0 \rangle \approx 
 - {1 \over N} {1 \over P_{RS}(M)} {dP_{RS}(M) \over dM} .
\end{equation}
Also, due to $\tilde{h} \ll 1$, Eq.~(\ref{defH33}) reduces to
\begin{equation}
 \langle s_0 \rangle = \tilde{h},
\end{equation}
and we get 
\begin{equation}
 \tilde{h} =
 - {1 \over N} {d \over dM} \log P_{RS}(M) .
\label{lnP_RS}
\end{equation}
Defining the effective potential $V_{\rm eff}^{(RS)}(M)$
(i.e., the Ginzburg -- Landau fixed-$M$ free energy) of the present
system (with defect) by
\begin{equation}
 P_{RS}(M) \propto \exp \{ - N V_{\rm eff}^{(RS)}(M) \} ,
\label{P_RS}
\end{equation}
we get immediately
\begin{equation}
 \tilde{h} = {d V_{\rm eff}^{(RS)}(M) \over dM} .
\label{dV_RS}
\end{equation}
For large systems, the relative contribution of the defect is small, and
thus $\tilde{h}(M)$ is well approximated by $V_{\rm eff}(M)$ for the
finite system {\em without} a defect:
\begin{equation}
 P(M)  \propto  \exp \{ - N V_{\rm eff}(M) \} ,
\end{equation}
\begin{equation}
 \tilde{h}  =  {d V_{\rm eff}(M) \over dM}
            +  \cdots
\label{hVeff}
\end{equation}
where the ellipsis stands for corrections vanishing at large $N$.

As is well known, for finite 3D Ising models in a cubic box with periodic
boundary conditions, the distribution $P(M)$ has a double-peak structure
\cite{Binder81,HilferW} at the critical point. Thus $V_{\rm eff}(M)$
has a double-well shape, which immediately explains why $\tilde{h}$
goes negative for small values of $M$. In Fig.~\ref{fig2} we show
the quantitative comparison of $\tilde{h}(M)$ (depicted by diamonds)
and $d V_{\rm eff} / dM$ (solid line). One observes that the 
correspondence between the points and the line clearly improves with
increasing lattice size. To extract $V_{\rm eff}$ from the 
Monte-Carlo-generated $P(M)$ (Fig.~\ref{fig1}) we have exploited
the fact, reported in \cite{TsyBlo}, that 
for the system under consideration,
$P(M)$ can be well approximated by the ansatz
\begin{equation}
 P(M) \propto \exp \Bigl\{  
 - \Bigl( {M^2 \over M_0^2} - 1 \Bigr)^2 
   \Bigl( a {M^2 \over M_0^2} + c \Bigr) 
 \Bigr\} , \label{PMansatz}
\end{equation}
which applies to the finite-size regime, i.e. the finite size is small
compared to the bulk correlation length. We have fitted the Monte
Carlo generated $P(M)$ data accordingly, determined the parameters
$a$ and $c$, and thus obtained $dV_{\rm eff}(M)/dM$ in a simple
polynomial form.

It is also worth mentioning that the shape of $P(M)$ for a given
geometry (in our case, a cubic box with periodic boundaries) is
universal at the critical point. That is, the parameters $a$ and $c$
have well-defined scaling limits when the system size grows to
infinity. These values,
$a = 0.158(2)$, $c = 0.776(2)$, have been determined in \cite{TsyBlo}
by making use of a special model in the 3D Ising universality class,
which has almost no corrections to scaling \cite{BLH}.
The corresponding scaling form of $dV_{\rm eff}(M)/dM$ is plotted
by the dashed line in Fig.~\ref{fig2}. One observes that deviations
from scaling (between the solid and the dashed lines) go down with
increasing size, as they should.

The results in Fig.~\ref{fig2} confirm the relation between the
observable $\tilde{h}(M)$ as defined above in the fixed-$M$ ensemble,
and the probability distribution $P(M)$ in the canonical ensemble.
The remaining discrepancy (between the diamonds and the solid line
in Fig.~\ref{fig2}) is due to the ``defect'' discussed above.
The question arises whether it is possible to modify our definition of
$\tilde{h}$ in order to suppress this discrepancy. We have found that 
this is indeed the case.  Up to this point,
we restricted ourselves to symmetric local environments (3+3$-$)
to define $\tilde{h}$ via Eq.~(\ref{defH33}). As has already been 
mentioned, using Eq.~(\ref{defH}) one may use other types of local
environments as well. In those cases the magnetization
$k \equiv \sum_{i=1}^6 s_i$ enters the definition:
\begin{equation}
 \langle s_0 \rangle = \tanh (\tilde{h} + \beta k) .
\end{equation}
Following the same arguments as before, we decompose the system in the
local spin $s_0$, its fixed neighbors, and the remaining system ($RS$).
This leads to the following generalization of Eq.~(\ref{RS1}):
\begin{eqnarray}
 \langle s_0 \rangle & = &
 {
 e^{ \beta k} P_{RS}(M-{k\over N}-{1\over N}) - 
 e^{-\beta k} P_{RS}(M-{k\over N}+{1\over N})
 \over
 e^{ \beta k} P_{RS}(M-{k\over N}-{1\over N}) + 
 e^{-\beta k} P_{RS}(M-{k\over N}+{1\over N})
 }      \nonumber \\
 & \approx &
 \tanh \beta k - {1\over \cosh^2 \beta k} {1\over N} 
   {1\over P_{RS}(M-{k\over N})}
   {dP_{RS}(M-{k\over N}) \over dM}  \nonumber \\
 & \approx &
 \tanh \Bigl( \beta k - {1\over N} {1\over P_{RS}(M)} 
       {dP_{RS}(M) \over dM} \Bigr).
\end{eqnarray}
Thus we arrive once again at Eqs.~(\ref{lnP_RS}-\ref{dV_RS}). But we
now have a different type of defect in the remaining system, and a shift
of $k/N$ in the magnetization of the remaining system; we neglect the
latter effect. Now it seems plausible that one can suppress the
influence of the defect by averaging over all configurations of the
defect, weighted with their natural occurrence probabilities.
Such an averaging should more faithfully reproduce the characteristics
of a system
{\em without} a defect. The modified determination of $\tilde{h}$
is as follows. Sample configurations from the fixed-$M$ ensemble.
For each spin determine its orientation ($+$ or $-$) and the type of
its local environment (type 0, \ldots, 6 for 0+6$-$, \ldots, 6+0$-$).
Accumulate these data by incrementing one out of 14 bins $N_{q,+}$,
$N_{q,-}$, where $q = 0 \ldots 6$ denotes the type of local environment,
and $+$ or $-$ denotes the local spin.
The resulting population numbers satisfy 
$\sum_{q=0}^6 ( N_{q,+} + N_{q,-} ) = N$. Then, for each $q$, find
$\langle s_0 \rangle_q = ( N_{q,+} - N_{q,-} )/( N_{q,+} + N_{q,-} )$
and compute $\tilde{h}_q$ according to Eq.~(\ref{defH}).
Finally,
\begin{equation}
 \tilde{h}_{\rm improved} = 
 {1\over N} \sum_{q=0}^6 \tilde{h}_q \cdot ( N_{q,+} + N_{q,-} ) .
\label{h_impr}
\end{equation}
Applying this definition to our simulation data, we observe that,
within the statistical accuracy, the discrepancy between $\tilde{h}(M)$
and $dV_{\rm eff}(M)/dM$ is indeed eliminated (Fig.~\ref{fig2}).

\section{Discussion and conclusions}
\label{sec:disc}
The relation (\ref{hVeff}) looks exactly the same as the standard 
relation between the field and magnetization in the canonical ensemble:
\begin{equation}
 h  =  {d \tilde{V}_{\rm eff}(M) \over dM} .
\label{hVeff_canon}
\end{equation}
The observed differences between the properties of $\tilde{h}(M)$
in the fixed-$M$ ensemble and $h(M)$ in the canonical ensemble,
the most prominent of which is the nonmonotonic behavior of $\tilde{h}(M)$
instead of the monotonic behavior of $h(M)$, can be traced to the different
definitions of the effective potential. The one that occurs in
Eq.~(\ref{hVeff}) is the fixed-$M$ free energy,
\begin{equation}
V_{\rm eff}(M)=-(1/N) \log Z_f(M) ,
\end{equation}
while the one that enters Eq.~(\ref{hVeff_canon})
is defined via a Legendre transformation:
\begin{equation}
\tilde{V}_{\rm eff}(M)=-(1/N) \log Z_c(h) + h M ,
\end{equation}
where
\begin{equation}
 M = \langle M \rangle_h
\end{equation}
is the canonical average of the magnetization in an external field $h$.
The partition functions $Z_f(M)$ and $Z_c(h)$ were defined in Section 
\ref{sec:intr}.
In a situation where fluctuations become negligible, the term $hM$ in
$\tilde{V}_{\rm eff}$ cancels the field dependence of the Boltzmann
weights. Then both definitions of the effective potential become 
equivalent, and both effective potentials approach the bulk form
so that the difference between $h$ and $\tilde{h}$ vanishes.

In a finite system, due to fluctuations, $\tilde{V}_{\rm eff}$
 differs from $V_{\rm eff}$.
For instance, at the Ising critical point, the double-well form of
$V_{\rm eff}$ is absent: $\tilde{V}_{\rm eff}$ has a single-well form.

Returning to Eq.~(\ref{dV_RS}), there is another finite-size effect:
the difference between $V_{\rm eff}^{(RS)}(M)$ and $V_{\rm eff}(M)$
due to the presence of the defect.
The relative contribution of the defect becomes small for large systems,
and it is further suppressed by the improved definition of $\tilde{h}$,
Eq.~(\ref{h_impr}).

Thus for large systems and sufficiently high $M$, 
when the correlation length is
small in comparison with the system size, the finite size effects
are suppressed, the difference between $V_{\rm eff}(M)$ 
and the bulk effective potential disappears, and our definition of 
$\tilde{h}(M)$ reproduces the expected bulk behavior.

\begin{figure}

\centerline{
\epsfxsize=12cm
\epsfbox{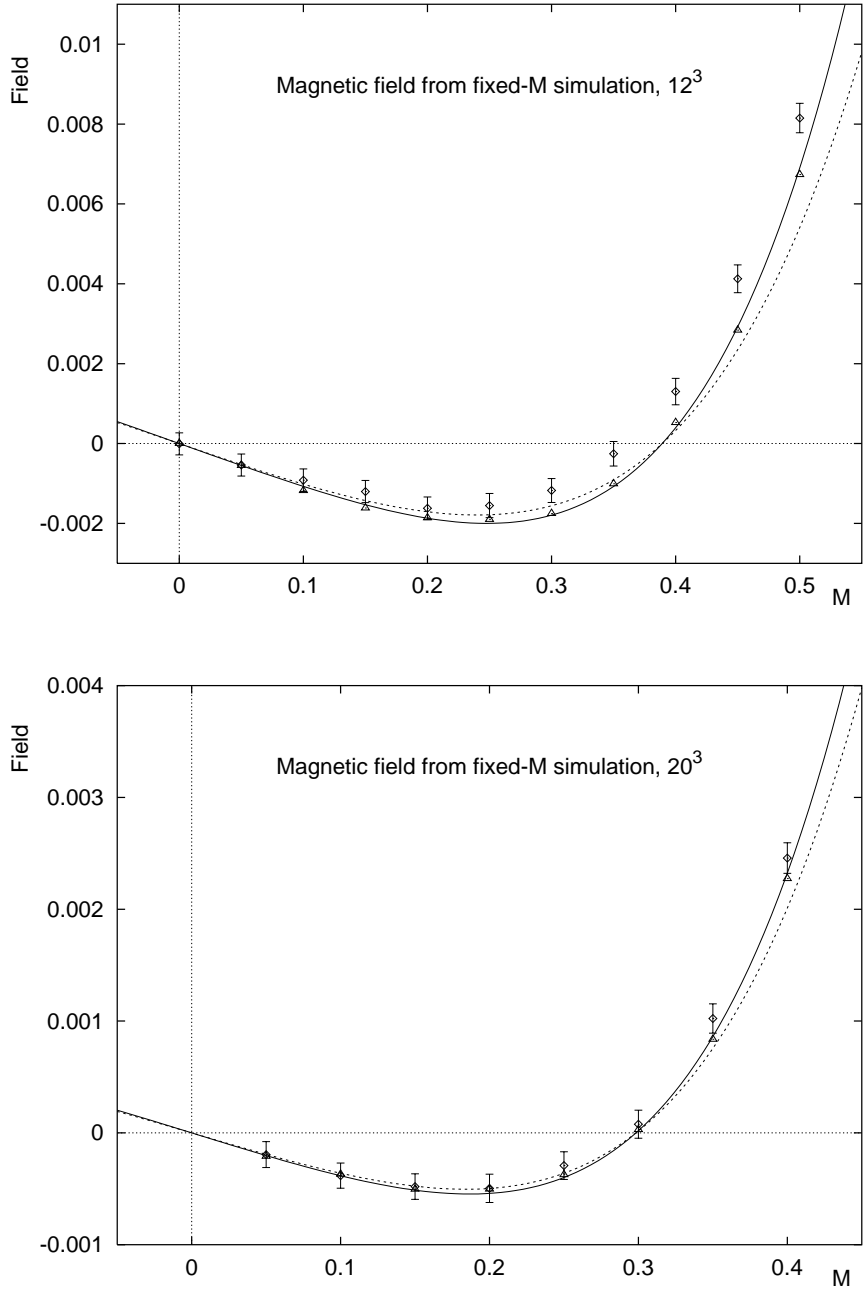}
\bigskip}

\caption{
Magnetic field $\tilde{h}$, computed as an observable in the fixed-$M$
ensemble Eq.~(\protect\ref{ZfixedM}). The temperature, boundary conditions
and lattice sizes are the same as in Fig.~\protect\ref{fig1}.
Results obtained from Eq.~(\protect\ref{defH33}), restricted to
spins with a local environment of the type 3+3$-$, are shown as diamonds.
Triangles correspond to the improved definition, Eq.~(\protect\ref{h_impr}).
Solid lines show $dV_{\rm eff}(M)/dM$, where $V_{\rm eff}(M)$ is exactly
the same as in Fig.~\protect\ref{fig1}. The dashed lines show the universal
shape of $dV_{\rm eff}(M)/dM$, using the universal (scaling-limit) values
of the parameters: $a = 0.158(2)$, $c = 0.776(2)$ \protect\cite{TsyBlo}
}
\label{fig2}
\end{figure}

In conclusion, we have studied the critical three-dimensional Ising model 
in the fixed-magnetization ensemble, in a cubic geometry with periodic
boundary conditions. This was done by means of the recently developed
geometric cluster Monte Carlo algorithm. We have defined a magnetic
field-like observable $\tilde{h}$ for this ensemble, studied its
dependence on the magnetization $M$ and explained its counter-intuitive
nonmonotonic behavior: $\tilde{h}$ first becomes negative and then
positive with increasing $M$ (Fig.~\ref{fig2}). We have provided a
quantitative description of $\tilde{h}(M)$ by establishing a close
relation with $P(M)$ --- the probability distribution of the magnetization
in the canonical ensemble. The nonmonotonic behavior of $\tilde{h}(M)$
can be understood as a manifestation of the same finite-size effect 
that is responsible for the double-peak shape of $P(M)$ at the 
critical point. Furthermore we have shown that, when fluctuations are
negligible, our definition reduces to the standard canonical relation
$M(h)$. Finally, we note that in the different context of
the simulation of a system of particles whose number is fixed, a
similar line of reasoning enables the determination of the chemical
potential of the particles \cite{Widom}.

\acknowledgements

We thank INTAS (grant CT93-0023) and DRSTP (Dutch Research School for
Theoretical Physics) for enabling one of us (M.T.) to visit Delft
University.

\end{document}